\documentclass[aps,twocolumn,prl,floatfix,showpacs]{revtex4}
\usepackage{graphicx}
\usepackage{amsfonts,amsmath}

\begin{document}
\title{Thermal non-equilibrium effects in quantum reflection}
\author{Viola Druzhinina$^1$}
%\email{viola.droujinina@physik.uni-freiburg.de}
\author{Marcel Mudrich$^1$}
\author{Florian Arnecke$^2$}
\author{Javier Madro{\~n}ero$^{2,3}$}
\author{Andreas Buchleitner$^1$}
\affiliation{$^1$Physikalisches Institut, Albert-Ludwigs Universit\"at Freiburg, D-79104 Freiburg, Germany}
\affiliation{$^2$Physik Department, Technische Universit{\"a}t M{\"u}nchen, D-85747 Garching, Germany}
\affiliation{$^3$Laboratoire de Physique Atomique, Mol\'eculaire et Optique (unit\'e PAMO), Universit\'e Catholique de Louvain, 2, chemin du Cyclotron, B-1348 Louvain-la-Neuve, Belgium}

\date{\today}
\begin{abstract}
We show that the quantum reflection coefficient of ultracold heavy atoms scattering off a dielectric surface can be tuned in a wide range by suitable choice of surface and environment temperatures. This effect results from a temperature dependent long-range repulsive part of the van der Waals-Casimir-Polder-Lifshitz atom-surface interaction potential.
\end{abstract}
\maketitle

The reflection of a matter wave from an attractive atom-surface interaction potential without reaching the classical turning point is known as quantum reflection (QR)~\cite{Pokrovskii}. The reflection probability approaches unity for vanishing primary kinetic (injection) energy of the atom, and can be enhanced by reducing the strength of the interaction potential. Furthermore, in thermal equilibrium light atoms are quantum reflected much more efficiently than heavy ones. These conditions have previously been realized in experiments in which neutral helium or hydrogen atoms were scattered off the surface of liquid helium~\cite{Masuhara,Walraven}. Later, neutral atoms prepared in the ground or in metastable states were reflected from a solid surface at grazing incidence~\cite{Shimizu1,Druzhinina}.

Rapid progress in laser cooling techniques, but also in surface preparation on the nanometer-scale, has made it possible to evidence low-velocity QR of Bose condensed sodium atoms at normal incidence with probabilities reaching 67\%, possibly limited by mean-field interactions of the Bose-Einstein condensate (BEC)~\cite{Pasquini2,Scott}. Such high values could only be achieved using a nanostructured silicon surface with effectively reduced surface density, thereby lowering the interaction potential. Tunable atom-surface interactions, using evanescent waves created by a laser beam to enhance QR, have also been studied~\cite{Cote1,Aspect2}. High reflection probabilities open the possibility to utilize QR for trapping purposes~\cite{Jurisch}.

In this Letter we present a new approach to controlling QR by exploiting thermal effects on the fundamental atom-surface interaction. We study QR of an atom from the surface of a dielectric solid body at thermal non-equilibrium. The Lifshitz force, which accounts for thermal fluctuations of the electromagnetic (EM) field, was found to add either attractively or repulsively to the long-range Casimir-Polder (CP) force, depending on whether the temperature of the substrate is higher or lower than that of the environment~\cite{Antezza}. This either leads to suppressed or to enhanced QR, respectively. Thus, a QR coefficient close to unity may be expected even for heavy rubidium $^{87}$Rb atoms at nano-Kelvin temperatures, scattering off an unstructured silicon (Si) surface, provided the environment is approximately 1000\,K hotter than the Si substrate.
The detailed unterstanding of temperature effects in QR is of considerable importance, not only from a fundamental point of view, but also in the context of experimental realizations, such as storage of ultracold atoms on miniaturized atom-optical devices ('atom-chips')~\cite{Folman}.

Let us start with an analysis of the force acting on a neutral atom near the surface of a dielectric body. It is caused by the interaction of the atom with the evanescent component of the thermal EM radiation field emitted by the body at temperature $T_S$. Additionally, the thermal blackbody radiation field reflected from the surface formes an intensity gradient causing a repulsive force acting between the atom and the surface~\cite{Antezza, Obrecht}. The temperature of the thermal blackbody radiation (\textit{e.\,g.}, from the vacuum chamber) defines the environment temperature and is denoted as $T_E$. All expressions for the interaction potentials employed in this Letter are valid as long as $T_E$ and $T_S$ stay well below the lowest electronic transition energy of the atom, which is of the order of $k_{\mathrm{B}}\times 10^4\ldots 10^5$\,K, such that absorption of thermal photons is suppressed.

For thermal equilibrium ($T_E=T_S=T$), the atom-surface interaction potential was calculated using the theory of thermal fluctuations~\cite{Lifshitz}, and can generally be expressed by~\cite{Antezza2}
\vspace{-0.3 cm}
\begin{equation}
\label{Eq:potentialtheq}
U_{\rm{th}}^{\rm{eq}}(r,T)=-\frac{k_B T\alpha_0}{4r^3}\frac{\varepsilon_0-1}{\varepsilon_0+1}G\left( \frac{r}{\lambda_T}\right).
\end{equation}
Here, $k_B$, $\lambda_T=\hbar c/ (k_B T)$ ($7.6~\mu$m at 300~K), $\alpha_0\equiv \alpha (0)$ and $\varepsilon_0\equiv \varepsilon(0)$ denote Boltzmann's constant, the thermal photon  wavelength, the static polarizability of the atom and the static dielectric permittivity of the surface, respectively. The function $G(r/\lambda_T)$ in Eq.~(\ref{Eq:potentialtheq}), which interpolats between the two asymptotic regions $r\ll \lambda_T$ and $r\gg \lambda_T$, is explicitly defined in Ref.~\cite{Antezza2}.

For $r\ll \lambda_T$, $G$ tends towards $3\phi (\epsilon_0)\lambda_T(\varepsilon_0+1)/ 2\pi r(\varepsilon_0-1)$.
The potential Eq.~(\ref{Eq:potentialtheq}) then adopts the well-known, temperature-independent CP form $U_{\rm{th}}^{\rm{eq}}(r)=-C_4/r^{4}$, with potential strength $C_4=3\alpha_0 \hbar c \phi (\varepsilon_0)/8\pi $~\cite{Casimir}, where the function $\phi (\varepsilon_0)$ describes the dielectric properties of the surface, and is defined in Ref.~\cite{Dz}. This potential is caused by vacuum fluctuations of the EM field and is dominant at distances $l\ll r\ll \lambda_T$, where $l=\lambda_{tr}/ 2\pi$ is the effective wavelength of electronic transitions of the atom. For $r\ll l \ll \lambda_T$, however, the non-retarded van der Waals (vdW) potential is dominant, which needs to be incorporated. With the subtitution $r/\lambda_T\rightarrow (r+l)/\lambda_T$ in function $G$, the potential Eq.~(\ref{Eq:potentialtheq}) approaches at $r\ll \lambda_T$ the general van der Waals-Casimir-Polder (vdWCP) form $-C_4 r^{-3}/(r+l)$~\cite{Shimizu1,Friedrich}, which includes both non-retarded vdW and retarded CP potentials.

In the opposite case, when $r\gg \lambda_T$, $G$ approaches unity, and the potential takes the form of the classical, temperature dependent Lifshitz potential, $U_{\rm{th}}^{\rm{eq}}(r,T)=-C_3(T)/r^3$, with strength $C_3 (T)=\alpha_0 k_{\mathrm{B}} T (\varepsilon_0-1)/4(\varepsilon_0+1)$. The Lifshitz potential has its origin in the thermal fluctuations of the EM field at finite temperature $T$~\cite{Lifshitz}.

If the system is out of thermal equilibrium, surface temperature
$T_S$ and environment temperature
$T_E$  differ from each other. $C_3(T)$ then turns into a function of $T_E$, and the Lifshitz potential aquires an additional term~\cite{Antezza,Henkel},
\vspace{-0.3 cm}
\begin{eqnarray}
\label{Eq:potentialthneq}
&  &U_{\rm{th}}^{\rm{neq}}(r,T_S,T_E)=-\frac{4\hbar \alpha_0}{\pi c^4 (\varepsilon_0-1)}\int_r^{\infty}dr' \int_0^{\infty} d\omega \times \nonumber\\
&  & \times \int_0^{\sqrt {\varepsilon_0-1} } dt\omega^4e^{-2r'\omega t/c} \left[ \frac{1}{e^{\frac{\omega\lambda _{T_S} }{c}}-1}-\frac{1}{e^{\frac{\omega\lambda _{T_E} }{c}}-1}\right] \times \nonumber\\
&  &\times t^2\sqrt{\varepsilon_0-1-t^2}\left[ 1+\frac{\varepsilon_0(2t^2+1)}{1+t^2(\varepsilon_0+1)}\right].
\end{eqnarray}
For distances $r\gg l$, we derived Eq.~(\ref{Eq:potentialthneq}) from the atom-surface force given in Ref.~\cite{Antezza}. At distances closer to the surface thermal fluctuations play no role, and $U_{\rm{th}}^{\rm{neq}}$ is negligible compared to the vdWCP potential. Asymptotically, for $r\gg \lambda_{T_S}/\sqrt{\varepsilon_0-1}$ and $r\gg \lambda_{T_E}/\sqrt{\varepsilon_0-1}$, the quantum mechanical $U_{\rm{th}}^{\rm{neq}}(r,T_S,T_E)$ behaves like $C_2(T_S,T_E)/r^2$ with potential strength $C_2(T_S,T_E)=\pi \alpha_0 k_{\mathrm{B}}^2 (T_E^2-T_S^2)(\varepsilon_0+1)/(12\hbar c\sqrt{\varepsilon_0-1})$~\cite{Antezza}. In this limit, $U_{\rm{th}}^{\rm{neq}}(r,T_S,T_E)$ dominates the Lifshitz eqilibrium potential $U_{\rm{th}}^{\rm{eq}}(r\gg \lambda_{T_E},T_E)\simeq -C_3(T_E)/r^3$, and therefore determines the asymptotic behavior of the full atom-dielectric surface interaction potential,
\begin{eqnarray}
\label{Eq:fullPot}
& & U_{\rm{th}}(r,T_S,T_E) = U_{\rm{th}}^{\rm{eq}}(r,T_E)+U_{\rm{th}}^{\rm{neq}}(r,T_S,T_E)\\
\label{Eq:C2}
& & \rightarrow \frac{C_2(T_S,T_E)}{r^2}, \, \, \, r\gg \left( \frac{\lambda_{T_S}}{\sqrt{\varepsilon_0-1}},\frac{\lambda_{T_E}}{\sqrt{\varepsilon_0-1}} \right).
\end{eqnarray}

An enhanced attractive potential at non-equilibrium with $T_S > T_E=310$~K was experimentally confirmed in~\cite{Obrecht},
%by measuring the collective oscillation frequency of the mechanical dipole mode of
for a $^{87}$Rb BEC placed $6\dots 11$~$\mu$m away from a fused-silica dielectric surface.
\begin{figure}[t]
\vspace{-0.8 cm}
\center
\center\resizebox{1.0\columnwidth}{!}{\includegraphics{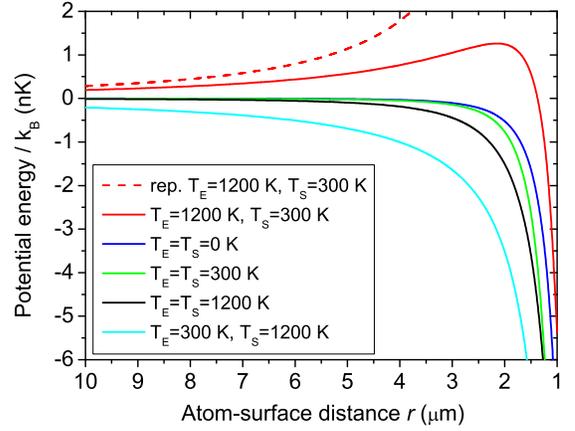}}
\vspace{-1 cm}
\caption{(Color online) Numerical vdWCP-Lifshitz full potential energy~(\ref{Eq:fullPot}), in units of temperature vs. distance between a $^{87}$Rb atom and a Si surface, for different environment ($T_E$) and surface ($T_S$) temperatures: $T_E=1200$~K, $T_S=300$~K,  $T_E=T_S=0$~K, $T_E=T_S=300$~K, and $T_E=300$~K, $T_S=1200$~K (solid lines from top to the bottom). The repulsive asymptote~(Eq.~(\ref{Eq:C2})) is represented by the dashed line.
 }
\vspace{-0.5 cm}
\label{fig:Potentials}
\end{figure}
In the case $T_E>T_S$, Eq.~(\ref{Eq:C2}) leads to a repulsive barrier in the full atom-surface potential (Eq.~(\ref{Eq:fullPot})).
%So far we know there is no experimental verification in this case.

For $^{87}$Rb atoms scattering off a Si surface ($\varepsilon_0\simeq12$, $l\simeq 130$ nm, $C_4\simeq7.6\times10^{-37}$eV/m$^4$) at variable $T_S$ and $T_E$, the numerically evaluated interaction potential curves~(Eq.~(\ref{Eq:fullPot})) are shown in Fig.~\ref{fig:Potentials}. At thermal equilibrium the potential is attractive. The curves corresponding to different temperatures $T_E=T_S$ (second, third and fourth solid lines from bottom to the top) clearly differ from each other. In the non-equilibrium case of a hot surface, $T_S>T_E$, the potential is strongly attractive (lowest solid line) in the entire range of $r$. In contrast, a hot environment ($T_E=1200$~K, $T_S=300$~K) induces a repulsive potential barrier (upper solid line). The potential with the repulsive barrier coincides with its asymptote~(\ref{Eq:C2}) (dashed line) at distances considerably larger than $\lambda_{T_E}/\sqrt{\varepsilon_0-1}\simeq 0.6~\mu$m and $\lambda_{T_S}/\sqrt{\varepsilon_0-1}\simeq 2.3~\mu$m. A hot environment at $T_E=1200$~K can easily be realized experimentally, \textit{e.\,g.}, by means of heated plates mounted in the vicinity of the dielectricum.

Since QR sensitively depends on the shape and magnitude of the interaction potential, we may expect a considerable influence of temperature on the QR coefficient. In order to evaluate this effect, we simulate the QR coefficient numerically by matching two WKB-wave functions, valid very far from and very close to the surface, with the exact numerical solution of the Schr{\"o}dinger equation, using the full numerical potential $U_{\rm{th}}(r,T_S,T_E)$ as input~\cite{Cote2}.
%In order to evaluate this effect, we simulate the QR coefficient numerically, by matching the WKB-wave function, which is valid very far from the surface, with the exact numerical solution of the Schr{\"o}dinger equation with potential of Eq.~(\ref{Eq:potential}) as input. As a boundary condition, the wave function has to be matched with the other WKB-wave function very close to the surface~\cite{Cote2}.
Possible classical reflection from the short-range repulsive wall or sticking to the surface by inelastic scattering are not taken into account.
%This reflection mechanism should be supressed in the experiment.

\begin{figure}[t]
\vspace{-1 cm}
\center
\center\resizebox{0.99\columnwidth}{!}{\includegraphics{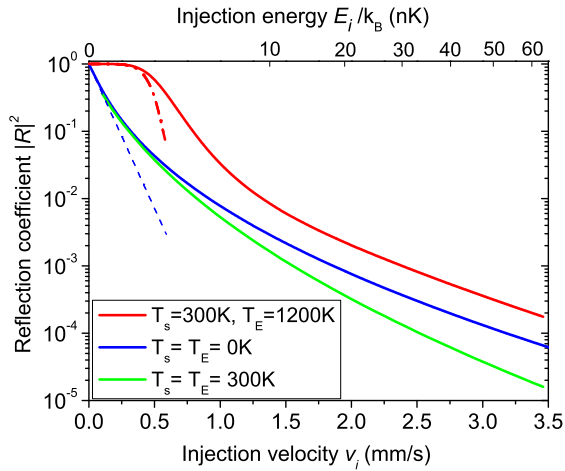}}
\vspace{-1 cm}
\caption{(Color online) Numerical QR coefficient of $^{87}$Rb from a Si surface, as a function of $v_i$ (bottom scale) and $E_i$, in units of temperature (upper scale) for different $T_E$ and $T_S$. The dashed and dash-dotted lines indicate the asymptotic behavior at low $v_i$ given by Eq.'s~(\ref{Eq:asimptota}) and~(\ref{Eq:asimptotanum}) with $\gamma=6.5$ and $b\simeq 2$~s/mm, respectively.}
\vspace{-0.2 cm}
\label{fig:Reflection}
\end{figure}

The numerically obtained QR coefficient of $^{87}$Rb scattering off a Si surface as a function of the atomic injection velocity perpendicular to the surface, $v_i$ ($=\hbar k_i/m$) (bottom scale), for different choices of $T_E$ and $T_S$ are depicted in Fig.~\ref{fig:Reflection}.  Note the top scale $E_i/k_B=mv_i^2/(2k_B)$. The middle solid curve represents QR from the vdWCP interaction potential, without taking temperature effects into account. The lower solid curve shows $|R|^2$ for the system at room temperature ($T_S=T_E=300$ K). Because of stronger attraction, the reflection probability is lowered. This might explain the measurement of smaller values of $|R|^{2}$ than the ones expected for single atom reflection~\cite{Pasquini}. The upper solid line in Fig.~\ref{fig:Reflection} displays QR evaluated using the full potential with a repulsive barrier with maximum $U_{\rm{bar}}=U(r_{\rm{bar}},T_S,T_E)\simeq 1.26$~nK$\times k_{\mathrm{B}}$ located at the distance $r=r_{\rm{bar}}\simeq 2~\mu$m. Clearly, $|R|^{2}$ is substantially increased. For $v_{i{\rm{bar}}}=\sqrt{2U_{\rm{bar}}/m}\simeq 0.49$~mm/s the atoms are $\simeq65$~$\%$ reflected, as opposed to $|R|^{2}\simeq5$~$\%$ of this value in thermal equilibrium at room temperature. But also at injection energies exceeding the barrier height, at which pure quantum reflection takes place, the reflection coefficient exceeds the one at thermal equilibrium conditions at room
temperature by at least one order of magnitude.

At zero-temperatures $T_E=T_S=0$, and at very low values of $v_i$ or $C_4$, the QR coefficient adopts the well-known exponential form (dashed line in Fig.~\ref{fig:Reflection})~\cite{Friedrich}
\begin{equation}
\label{Eq:asimptota}
|R|^{2}  \simeq \exp[- 4 k_{i} \beta _4]=\exp[-\frac{4m\beta_4}{\hbar}v_i] ,\, \, \, \, \, k_i\beta_4 \ll 1.
\end{equation}
Here, $C_4$ enters in terms of the length parameter $\beta _4 =(2m C_4/\hbar^2)^{1/2}$.

At non-equilibrium, for $v_i \rightarrow 0$, analysing the numerical results for $|R|^2$ reveals the exponential asymptote (dash-dotted line in Fig.~\ref{fig:Reflection})
\begin{equation}
\label{Eq:asimptotanum}
|R|^{2} \simeq \exp[- (bv_i)^{\gamma}].
\end{equation}
Our analytic investigations based on the theory developed in~\cite{Arnecke} allow us to define the parameter $\gamma$ as $\gamma=\sqrt{1+4\beta_0}$. This parameter differs from unity if the system is out of thermal equilibrium through its dependence on $\beta_0=2m C_2(T_E,T_S)/ \hbar ^2$.
%For increasing $\gamma$ (\textit{e.\,g.}, for heavy atoms), $|R|^2$ approaches unity more rapidly for decreasing $v_i$.
Parameter $b$ in Eq.~(\ref{Eq:asimptotanum}) unfortunately cannot be represented by a simple analytic expression. The dash-dotted curve in Fig.~\ref{fig:Reflection} is obtained by fitting expression~(\ref{Eq:asimptotanum}) to our numerical results in the region of small $v_i$ with the fit parameter $b\simeq 2$~s/mm.
%with some parameter $b$ (determined by the offset of the straight line, $\ln(b)=\ln(-\ln(|R|^2))$ at $\ln(v_i)=0$) At $v_i\rightarrow 0$ we can predict the same result by comparing the numerically obtained $|R|^2$ with the reflection probability from some potential $U(r)=-C_{\alpha}/r^{\alpha}+C_2/r^2$, with $\alpha>2$, analytically calculated using Jost functions. To this end we determine the leading term of the scattering phase $\delta_{\gamma}$ of the reflected wave.  The QR coefficient can be represented as $|R|^{2} \simeq 1-4\times \,\it{Im} (\tan \delta_{\gamma})$, and is approximately given by Eq.~(\ref{Eq:asimptotanum}) with parameter $b$,
%\begin{equation}
%\label{Eq:b}
%b \simeq -\frac{\pi \Gamma[-\frac{\gamma}{\alpha-2}]\sin (\frac{\pi \gamma}{\alpha-2})(\frac{m}{\hbar})^{\gamma}(\frac{2m}{\hbar^2}C_{\alpha})^{\frac{\gamma}{\alpha-2}}}{2^{\gamma-2}(\alpha-2)^{\frac{2\gamma}{\alpha-2}}\Gamma[\frac{\gamma}{2}]\Gamma[1+\frac{\gamma}{2}]\Gamma[\frac{\gamma}{\alpha-2}]}.
%\end{equation}

QR probabilities for various atomic species with strongly differing properties are best compared in terms of the dimensionless parameter $k_i \beta_4$. For this comparison we consider three different atomic species with pairwise similar $m$ or $\alpha_0{\--}$ $^{87}$Rb,  $^{4}$He$^*$ in the metastable triplet state, and ground state $^4$He${\--}$ scattering off the same Si surface. Temperatures are held fixed at $T_E=1200$~K and $T_S=300$~K.
\begin{figure}[t]
\vspace{-1 cm}
\center
\center\resizebox{1.0\columnwidth}{!}{\includegraphics{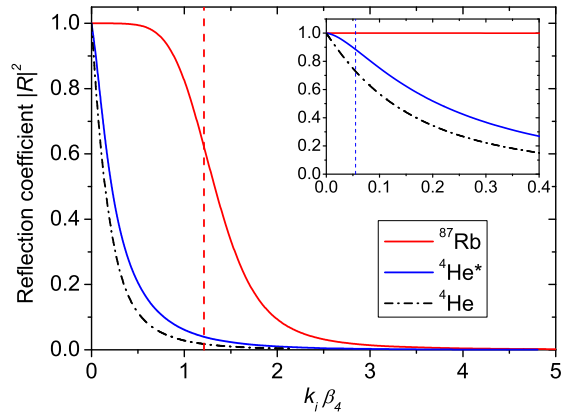}}
\vspace{-1 cm}
\caption{(Color online) Numerical QR coefficient, as a function of the dimensionless parameter $k_{i} \beta_4$, for different atomic species scattering off a Si surface, at $T_E=1200$~K and $T_S=300$~K. Inset: Zooms into the region of small $k_i \beta_4$-parameter. The vertical dashed lines denote the values of $(k_i \beta_4)_{\rm{bar}}$, at which $E_i=U_{\rm{bar}}$.}
\vspace{-0.5 cm}
\label{fig:SShape}
\end{figure}
 %~%~\cite{Zimmermann},\cite{ShimizuMetastabHe}\cite{Thomas})
The numerically obtained reflection coefficients are displayed in Fig.~\ref{fig:SShape}. Given the atomic species and temperatures, the parameter $\gamma$, which determines the asymptotic behavior, amounts to $6.5$ ($^{87}$Rb: upper solid line), $1.7$ ($^4$He$^*$: lower solid line) and $1.004$ ($^4$He: dash-dotted line). The inset zooms into the same data at small values of $k_i \beta_4$. At injection energies equal to the height of the repulsive barrier, $k_i \beta_4$ has values $(k_i \beta_4)_{\rm{bar}}=1.21$, $0.055$, and $2.4\times10^{-4}$, for Rb, He$^*$, and He, respectively. For Rb and He$^*$, the values for $(k_i \beta_4)_{\rm{bar}}$ are depicted as vertical dashed lines in Fig.~\ref{fig:SShape} and in its inset, respectively. They delimit the region of pure above-barrier reflection, $E_i> U_{\rm{bar}}$, from the one where matter waves are reflected from the classical turning point, $E_i<U_{\rm{bar}}$. For classical particles, $|R|^2(k_i\beta_4)$ would exhibit a Heavyside step function at $(k_i\beta_4)_{\rm{bar}}$. The S-shaped behavior of $|R|^2$ highlights the quantum nature of the reflection process from a potential with a repulsive barrier. This behavior  is modified by tunneling ($k_i\beta_4<(k_i \beta_4)_{\rm{bar}}$), and by above-barrier reflection ($k_i\beta_4>(k_i \beta_4)_{\rm{bar}}$), respectively.

Since $^{87}$Rb and He$^{*}$ have similar static polarizabilities ($\alpha_0= 47.25$~$\mathrm{\AA}^3$ for $^{87}$Rb, and $46.8$~$ \mathrm{\AA}^3$ for $^{4}$He$^*$), their atom-surface potentials nearly coincide, with barrier heights corresponding to a temperature $T_{\rm{bar}}=\hbar ^2(k_i \beta_4)_{\rm{bar}}^2/(2mk_{\mathrm{B}} \beta_4^2)\simeq 1.26$~nK. Fig.~\ref{fig:SShape} shows, however, that the QR coefficients at $k_i\beta_4=(k_i \beta_4)_{\rm{bar}}$ (the intersections of dashed lines with solid ones) are quite different for these two species. At the same fixed $E_i$, light atoms are reflected much more efficiently than heavy ones. This mass dependence is a typical feature, which also applies to above-barrier reflection from a pure attractive potential. Without a repulsive barrier, the upper solid curve ($^{87}$Rb) would be shifted to even smaller values of $k_i\beta_4$ than the dash-dotted one ($^4$He, $\alpha_0= 0.205$~$ \mathrm{\AA}^3$). In contrast, in the presence of a barrier, $|R|^2$ increases significantly for heavy $^{87}$Rb. Furthermore, for $^4$He and $^4$He$^{*}$ having same $m$ but different $\alpha_0$, the QR coefficients have different magnitudes at small fixed value of $k_i\beta_4$, due to different values of $\gamma$.
\begin{figure}[t]
\vspace{-0.8 cm}
\center
\center\resizebox{1.0\columnwidth}{!}{\includegraphics{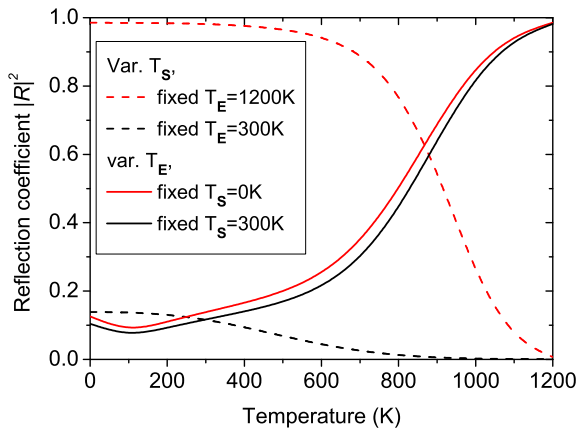}}
\vspace{-1 cm}
\caption{(Color online) Numerical QR coefficient of $^{87}$Rb atoms scattering off a Si surface, as a function of environment or surface temperature, with the other one fixe. Upper and lower dashed lines: Dependence on $T_S$, at fixed $T_E=1200$~K and $T_E=300$~K, respectively. Upper and lower solid lines: Dependence on $T_E$, at fixed $T_S=0$~K and $T_S=300$~K, respectively. $k_i \beta_4$ was set to $0.68$.}
\vspace{-0.5 cm}
\label{fig:TDependence}
\end{figure}

In order to illustrate the potential to control QR efficiency by adjusting the system temperatures, Fig.~\ref{fig:TDependence} shows
the dependence of $|R|^2$ on either $T_E$ or $T_S$, with the other one fixed, for the $^{87}$Rb/Si system. We choose $k_i \beta_4=0.68$, which corresponds to $E_i\simeq 0.4$~nK$\times k_{\mathrm{B}}$. At $T_S=T_E$ all curves in Fig.~\ref{fig:TDependence} have very small but distinct values, due to above-barrier reflection from the attractive potential in thermal equilibrium at different temperatures. At $T_E$ between $0$ and $200$~K the solid curves reveal the competition between $U_{\rm{th}}^{\rm{eq}}(r,T_E)$ and $U_{\rm{th}}^{\rm{neq}}(r,T_S,T_E)$. As $T_S$ falls below the fixed value of $T_E$ (dashed curves) or $T_E$ rises above the fixed value of $T_S$ (solid curves), a repulsive barrier emerges in the atom-surface potential. Consequently, the QR coefficient grows nearly to unity, except for the $T_S$-dependence at $T_E=300$~K, at which the barrier height remains very low. Clearly, heating the environment to a temperature which is by up to $900$~K higher than the one of the surface, drastically enhances the QR probability in the entire temperature range, as opposed to a weak increase of $|R|^2$ when just cooling the surface close to $0$~K.

In conclusion, we have shown that the QR probability is significantly enhanced in the presence of a repulsive barrier in the atom-surface interaction potential, which emerges when the environment temperature exceeds the one of the surface.
%In the limit $v_i\rightarrow 0$ the QR coefficient can be analyzed analytically in terms of atomic and surface properties, as well as of the environment temperature.
By changing one of the temperatures it is possible to vary the QR probability in a wide range, in particular when using heavy atoms. In analogy to a macroscopic sphere near a surface~\cite{Munday}, a heavy ultra-cold atom could be quantum levitated a few micrometers above a surface. This opens new perspectives for guiding and trapping ultracold atoms on surfaces, \textit{e.\,g.} on atom chips.

\acknowledgments{
Helpful discussions with M. Antezza, H. Friedrich and T. Wellens are acknowledged.
We are grateful to the Schlieben-Lange-Program fellowship for supporting the work of V.D.}

\end{document}